\documentclass[aps,twocolumn]{revtex4}
\usepackage{graphicx}
\begin{document}
\title{Vanishing Thermal Mass in the Strongly Coupled QCD/QED Medium}
\author{Hisao Nakkagawa}%
 \email{nakk@daibutsu.nara-u.ac.jp}
\author{Hiroshi Yokota}%
 \email{yokotah@daibutsu.nara-u.ac.jp}
\author{Koji Yoshida}%
 \email{yoshidak@daibutsu.nara-u.ac.jp}
\affiliation{%
 Institute for Natural Science, Nara University, 1500 Misasagi-cho, Nara 631-8502, Japan
}%

\date{\today}
\begin{abstract}
In this paper we perform a nonperturbative analysis of a thermal quasifermion in thermal QCD/QED by studying its self-energy 
function through the Dyson-Schwinger equation with the hard-thermal-loop resummed improved ladder kernel. Our 
analysis reveals several interesting results, some of which may force us to change the image of the thermal 
quasifermion: (1) The thermal mass of a quasifermion begins to decrease as the coupling gets stronger and finally disappears in the strong coupling region, 
(2) the imaginary part of the chiral invariant mass function (i.e., the decay width 
of the quasifermion) persists to have $O(g^2 T \log (1/g))$ behavior. Present results suggest 
that in the recently produced strongly coupled quark-gluon-plasma, the thermal mass of a quasifermion should vanish. 
We also briefly comment on evidence of the existence of a massless, or an ultrasoft mode.
\end{abstract}
\maketitle

\section{Introduction} 
High energy heavy ion collision experiments carried out at the BNL Relativistic Heavy Ion Collider (RHIC) and CERN LHC 
have achieved
the creation of a quark-gluon plasma (QGP) phase, thus liberating the quark and gluon degrees of freedom. 
Unexpectedly, however, the produced QGP medium showed the property close to that of a perfect fluid. 
This fact leads us to the understanding that the QGP produced in the energy-region of the BNL RHIC 
is a strongly interacting system of quarks and gluons, namely, the strongly coupled QGP (sQGP)~\cite{review}.
With this finding, properties of a thermal quasi-particle in the QGP phase attracted our interest once again.

Up to now, most of the theoretical findings on thermal quasifermion in the QGP are obtained through the analyses 
with the assumption of weakly coupled QGP at high temperature, i.e., analyses through the hard-thermal-loop (HTL) 
resummed effective perturbation calculation\cite{Rebhan}, or those through the one-loop calculation by 
replacing the thermal gluon with the massive vector boson~\cite{Kitazawa}. Such analyses, however, cannot be 
justified in studying the thermal quasiparticle in the sQGP created in the energy region of BNL RHIC, 
requiring us to perform nonperturbative studies. Calculations of correlators within lattice QCD are performed 
in Euclidean space and give interesting results~\cite{Schaefer}. However, strictly speaking it is not possible to 
carry out an analytic continuation that is necessary to determine the spectral function. In addition it is difficult 
on the lattice to respect the chiral symmetry that should be restored in the sQGP phase, though we are interested 
in the property of thermal quasiparticle in the chiral symmetric sQGP phase.

In this paper we perform a nonperturbative analysis of thermal quasifermion in a thermal QCD/QED by studying its self-energy 
function through the Dyson-Schwinger equation (DSE) with the HTL resummed improved ladder kernel. Our 
analysis may overcome the problems in the previous analyses listed above for the following reasons: (1) it is a nonperturbative QCD/QED 
analysis, (2) we study the DSE in the real-time formalism of thermal field theory, which is suitable for
the direct calculation of the propagator, or the spectral function, (3) we use the HTL resummed thermal 
gauge boson (gluon/photon) 
propagator as an interaction kernel of the DSE, and take into account the quasiparticle decay processes 
by accurately studying the imaginary part of the self-energy function, and finally, (4) we present an analysis 
based on the DSE that respects the chiral symmetry and describes its dynamical breaking and restoration. Our 
analysis is nothing but an application of our formalism employing the DSE to the study of thermal 
quasi-femion on the strongly coupled QCD/QED medium with chiral symmetry~\cite{Nakk}.

Analogous studies employing the DSE are carried out by several groups~\cite{Harada}. All these analyses 
solve the DSE in the imaginary-time formalism, and try to perform an analytic continuation. Harada et al. 
study the DSE with a ladder kernel in which the vertex function and gauge boson propagator are replaced with 
the tree level quantities, while Qin et al. and Mueller et al. use the maximum entropy method to compute 
the quark spectral density.

Our analysis reveals several interesting results, some of which may force us to change the image of thermal 
quasifermion: (1) While the thermal mass of a quasifermion begins to decrease, the strength of the coupling gets stronger 
and finally disappears in the strong coupling region, thus showing a property of a massless 
particle, (2) its imaginary part (i.e., the decay width) persists to have $O(g^2T \log (1/g))$ behavior. These results 
suggest that in the recently produced strongly coupled QGP, the thermal mass of a quasifermion should vanish.

In the present paper, we report the results of the analysis by focusing mainly on the behavior of chiral invariant 
thermal mass and its imaginary part in the strongly coupled gauge theories. Full results on properties of the 
thermal quasifermion will be given in a separate paper~\cite{Nakk2}.

\section{HTL resummed improved ladder DSE for quasifermion self-energy function $\Sigma_R$} 
The retarded quasifermion propagator $S_R(P), P=(p_0, \textrm{\bf p})$, is expressed by
\begin{equation}
   S_R(P) = \frac{1}{P\!\!\!\!/ + i \epsilon \gamma^0 - \Sigma_R (P)} ,
\end{equation}
where $\Sigma_R$ is the quasifermion self-energy function that is tensor decomposed in a chiral
symmetric phase at finite temperature as follows:
\begin{equation}
   \Sigma_R(P) = (1-A(P)) p_i \gamma^i - B(P) \gamma^0 .
\end{equation}
$A(P)$ is the inverse of the fermion wave function renormalization function, and $B(P)$ is the 
chiral invariant mass function. The c-number mass function does not appear in the chiral symmetric phase.

As for the interaction kernel of the DSE, we use the tree vertex and the HTL resummed gauge boson 
propagator, and get the HTL resummed improved ladder DSE to determine the scalar invariants $A(P)$ and $B(P)$. 
(N.B.: To the longitudinal gauge boson propagator we apply the instantaneous exchange approximation, i.e., 
the zero-th component of the longitudinal gauge boson momentum $q_0$ is set to zero.) In this paper we study 
the massless QCD/QED in the Landau gauge, and the DSE to determine $A$ and $B$ becomes~\cite{foot}
\begin{eqnarray}
\!\!\!\! & &- i \Sigma_R(P) = - \frac{g^2}{2} \int \frac{d^4K}{(2 \pi)^4} \nonumber \\
 & & \ \ \times \left[ {}^* \Gamma^{\mu}_{RAA}(-P,K,P-K) S_{RA}(-K,K) \right. \nonumber \\
 & & \ \ \ \ \times {}^* \Gamma^{\nu}_{RAA} (-K,P,K-P) {}^*G_{RR,\mu \nu}(K-P,P-K)
             \nonumber \\
 & & \ \  + {}^* \Gamma^{\mu}_{RAA}(-P,K,P-K) S_{RR}(-K,K) \nonumber \\
 & & \ \ \ \ \left. \times {}^* \Gamma^{\nu}_{AAR} (-K,P,K-P) {}^*G_{RA,\mu \nu}(K-P,P-K) 
              \right] , \nonumber \\
\end{eqnarray}
where $^*G_{\mu\nu}$ is the HTL resummed gauge boson propagator~\cite{Klimov} and $^*\Gamma_{\mu} = \gamma_{\mu}$ 
in the present approximation.

The HTL resummed effective perturbation analysis enables us to study thermal physics of $O(gT)$. 
Thus, we expect that the nonperturbative analysis employing the DSE with the HTL resummed dressed 
kernel at least takes into account the important effects of thermal fluctuations up to $O(gT)$, 
and enables us to give reliable solutions over wider range of couplings and temperatures, i.e., 
the region $\hspace{0.3em}\raisebox{0.4ex}{$<$}\hspace{-0.75em}\raisebox{-.7ex}{$\sim$}\hspace{0.3em} O(gT)$.

In the chiral symmetric phase the fermion propagator can be expressed as 
\begin{equation}
S_R (P) = \frac12 \left[ \frac{1}{D_+} \left( \gamma^0 + \frac{p_i \gamma^i}{p} \right)
          + \frac{1}{D_-} \left( \gamma^0 - \frac{p_i \gamma^i}{p} \right) \right] 
\end{equation}
where
\begin{equation}
        D_{\pm}(P)=p_0+B(p_0,p) \mp p A(p_0,p)
\end{equation}
with
\begin{eqnarray}
             \mbox{Re} [D_+(p_0, p)] &=&  - \mbox{Re} [D_-(-p_0, p)], \\
             \mbox{Im} [D_+(p_0, p)] &=&   \mbox{Im} [D_-(-p_0, p)].
\end{eqnarray}

\section{Quasifermion dispersion law} \ \ 
The quasifermion pole is defined by the zero-point of the inverse of the chiral invariant fermion 
propagator $D_{\pm}(P) \equiv  D_{\pm}(p_0, p)$,
\begin{equation}
        \mbox{Re} [D_{\pm}(p_0 = \omega_{\pm}, p)] = 0,
\end{equation}
which determines the dispersion law of this pole. In Fig.~1 we give the quasifermion dispersion law $\omega=\omega_{\pm}(p)$ 
at small coupling $\alpha \equiv g^2/4\pi$ and at moderately high temperature~\cite{foot2}.
It should be noted that, as can be seen in Fig.~1, in the region of weak coupling 
strength $\alpha \hspace{0.3em}\raisebox{0.4ex}{$<$}\hspace{-0.75em}\raisebox{-.7ex}{$\sim$}\hspace{0.3em} 0.01$,
the dispersion law lies on a universal curve determined by the HTL calculations~\cite{Klimov}. 
Thus, the result shows a good agreement with the HTL resummed 
effective perturbation calculation. The quasifermion has a definite thermal mass of $O(gT)$, 
and the collective plasmino mode exhibits a minimum at $p \neq 0$ and vanishes rapidly on to the light cone.
\begin{figure}[htbp]
  \vspace*{-0.5cm}
  \centerline{\includegraphics[width=7cm,clip]{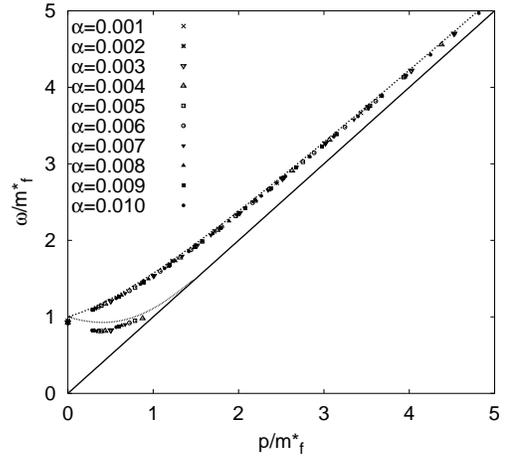}} \vspace{-0.4cm}
  \caption{Quasifermion dispersion law at small coupling and at moderately high temperature $T=0.3$ (see the text
  for details). The dotted curves are the dispersion law of the quasifermion and of the plasmino determined
  through the HTL calculation.}
\end{figure}

Two comments should be added.
\begin{description}
\vspace{-0.2cm}
\item{1)} \ \ In this region of temperature and coupling constant, the dispersion law determined by the 
zero-point of $D_+$ agrees well and almost coincides with that determined by the peak of the spectral 
density $\rho_+$. The spectral function of quasifermion $\rho_{\pm}$ is defined by
\begin{equation}
  \rho_{\pm}(P) = - \frac{1}{\pi} \mbox{Im} \frac{1}{D_{\pm}(P)} .
\end{equation}
\item{2)} \ \ Figure 1 shows that the quasifermion energy $\omega_+(p)$ approaches $m^*_f$ as 
$p \to 0$: namely, the thermal mass of the quasifermion is $m^*_f$
\begin{eqnarray}
     \left( \frac{m^*_f}{m_f} \right)^2 &=& 1 - \frac{4 g}{\pi} \left[
            - \frac{g}{2 \pi} + \sqrt{ \frac{g^2}{4\pi^2} + \frac13} \right] , \\
         m_f^2  &\equiv& \frac{g^2T^2}{8} \nonumber
\end{eqnarray}
which is determined through the next-to-leading order calculation of HTL resummed effective perturbation 
theory~\cite{Rebhan,Rebhan2}.
\end{description}

\section{Vanishing thermal mass in the strongly coupled QCD/QED medium} 
Now let us study how the result shown in Fig.~1 changes as the coupling gets stronger, namely, in the region of intermediate 
to strong couplings. First let us see the quasifermion dispersion law in the small momentum region. (N.B. :
Temperatures and couplings that we are studying belong to the chiral symmetric phase~\cite{foot2}.) 
\begin{figure}[htbp] 
    \vspace*{-0.5cm}
    \centerline{\includegraphics[width=7cm]{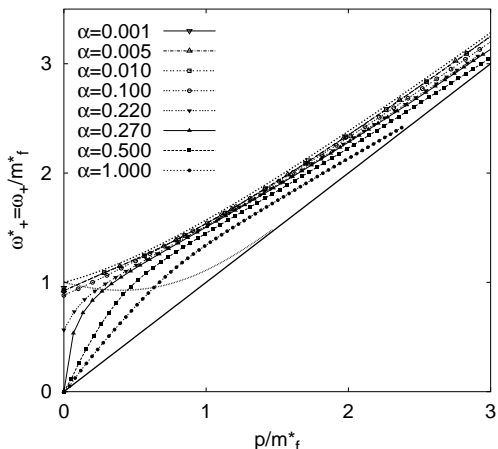}} \vspace{-0.4cm}
    \caption{Quasifermion dispersion law at $T=0.3$ for a range of couplings from weak to
     strong couplings (see the text for details). For simplicity, we show only the fermion branch.}
\end{figure}

Figure 2 shows the coupling $\alpha$ dependence of the normalized fermion dispersion law at $T=0.3$ as the 
coupling $\alpha$ becomes stronger, where the normalization scale is the next-to-leading order thermal 
mass $m^*_f$. For simplicity, in Fig.~2 we show only the fermion branch. Though in the weak coupling 
region we get the solution in good agreement with the HTL resummed perturbation analyses, as the 
coupling becomes stronger from the intermediate to the strong coupling region, the normalized thermal mass 
$\omega^*_+(p=0) \equiv \omega_+(p=0)/m^*_f$ begins to decrease from 1 and finally tends to zero 
(in the region $\alpha \hspace{0.3em}\raisebox{0.4ex}{$>$}\hspace{-0.75em}\raisebox{-.7ex}{$\sim$}\hspace{0.3em} 0.27$ in Fig.~2).
Namely, in the thermal QCD/QED medium, the thermal mass of the quasifermion begins to decrease as the strength of the coupling 
gets stronger and finally disappears in the strong coupling region.
This fact suggests that in the recently produced strongly coupled QGP, the thermal mass of the quasifermion should vanish or at 
least become significantly lighter compared to the value in the weakly coupled QGP. 
\begin{figure}[htbp] 
\vspace*{-0.5cm}
\centerline{\includegraphics[width=7.5cm]{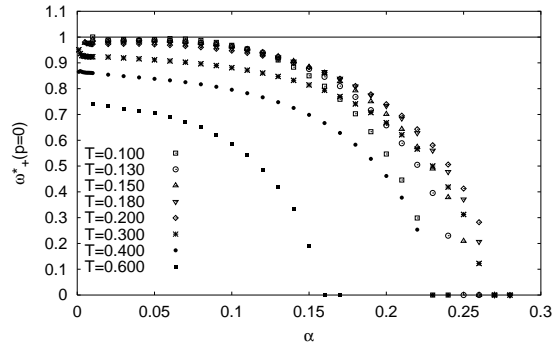}} \vspace{-0.4cm}
\caption{The $\alpha$ dependence of the normalized thermal mass $\omega^*_+(p=0)$ (see the text for details).}
\end{figure}

To see the above behavior of the thermal mass more clearly, in Fig.~3 we show the normalized thermal 
mass $\omega^*_+(p=0)$ as a function of $\alpha$. In the small coupling 
region ($\alpha \hspace{0.3em}\raisebox{0.4ex}{$<$}\hspace{-0.75em}\raisebox{-.7ex}{$\sim$}\hspace{0.3em} 0.1$) around 
the temperature range $T=0.1 \sim 0.2$, results of the thermal mass agree well with those of the HTL resummed 
perturbation calculation. The thermal mass $\omega_+(p=0)$ decreases and finally 
vanishes as the coupling gets stronger from the intermediate to the strong coupling region.
Analogous behavior of the thermal mass $\omega_+(p=0)$ appears in the temperature
dependence~\cite{foot3,foot4}.
\begin{figure}[htbp] 
   \vspace*{-0.5cm}
   \centerline{\includegraphics[width=7.5cm]{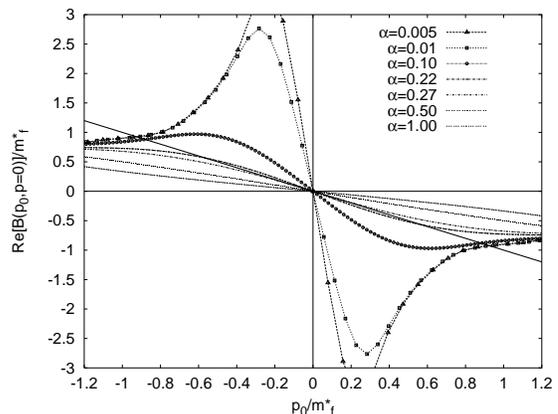}} \vspace{-0.4cm}
   \caption{The $p_0$ dependence of Re[$B(p_0, p=0)$] at $T=0.3$.}
\end{figure}

This behavior of the thermal mass is determined by the behavior of the chiral invariant mass function Re[$B(p_0, p)$]. 
In Fig.~4 we show the $p_0$ dependence of Re[$B(p_0, p=0)$] at $T=0.3$. At small coupling Re[$B(p_0, p=0)$] 
has a steep valley/peak structure in the small $p_0$ region, but as the coupling becomes stronger this structure eventually
disappears and Re[$B(p_0, p=0)$] begins to behave almost as a straight line. 

The thermal mass is given by the solution of Re[$B(p_0,p=0)$]=$-p_0$, which is the $p_0$ coordinate of the intersection 
point of the drawn curve of Re[$B(p_0,p=0)$] and the straight solid line through the origin with a slope -1 in Fig.~4.
At first we can see with this figure that at 
small coupling there are three intersection points: the one with positive $p_0$, one with negative $p_0$, and 
one at $p_0=0$, which corresponds to the quasifermion, the plasmino, and the massless (or, ultrasoft) 
modes~\cite{Kitazawa,Hidaka}, respectively. 
As the coupling becomes stronger ($\alpha \hspace{0.3em}\raisebox{0.4ex}{$>$}\hspace{-0.75em}\raisebox{-.7ex}{$\sim$}\hspace{0.3em} 0.27$ at $T=0.3$), 
however, there is only one intersection point at $p_0=0$, 
representing the massless pole in the fermion propagator. 
Thus we can understand the behavior in Fig.~3; namely, 
$\omega^*_+(p=0) \equiv \omega_+(p=0)/m^*_f$ is unity in the weak coupling region, and zero in the strong 
coupling region ($\alpha \hspace{0.3em}\raisebox{0.4ex}{$>$}\hspace{-0.75em}\raisebox{-.7ex}{$\sim$}\hspace{0.3em} 0.27$ at $T=0.3$); 
therefore, the fermion thermal mass vanishes completely in the corresponding strong coupling region.

\section{The third peak, or the massless mode} \ \ The quasifermion and the plasmino modes are well understood 
in the HTL resummed analyses, the latter being the collective mode to appear in the thermal environment. 
What is the third mode? Is it nothing but evidence of the massless, or the ultrasoft mode? Is there any signature in our analysis? 

To clarify this question, in Fig.~5(a) we give the spectral density 
$\rho_+(p_0,p=0)$ in the small coupling region ($\alpha=0.001, T=0.4$). 
Two sharp peaks, representing the quasifermion and the plasmino poles, are clearly seen, 
and the existence of a slight ``peak'' can also be recognized around $p_0 = 0$. 
To see this more clearly, in Fig.~5b we show the rescaled version of 
Fig.~5a, where we can clearly see the peak structure around $p_0=0$. This third peak is 
nothing but convincing evidence of the existence of a massless or an ultrasoft mode in the weak coupling region~\cite{Kitazawa,Hidaka}.
This peak is indistinctively slight compared to the sharp quasifermion/plasmino peak. This problem will
be fully discussed in a separate paper~\cite{Nakk2}.
\begin{figure}[htbp] 
  \vspace*{-0.5cm}
  \centerline{\includegraphics[width=7.5cm]{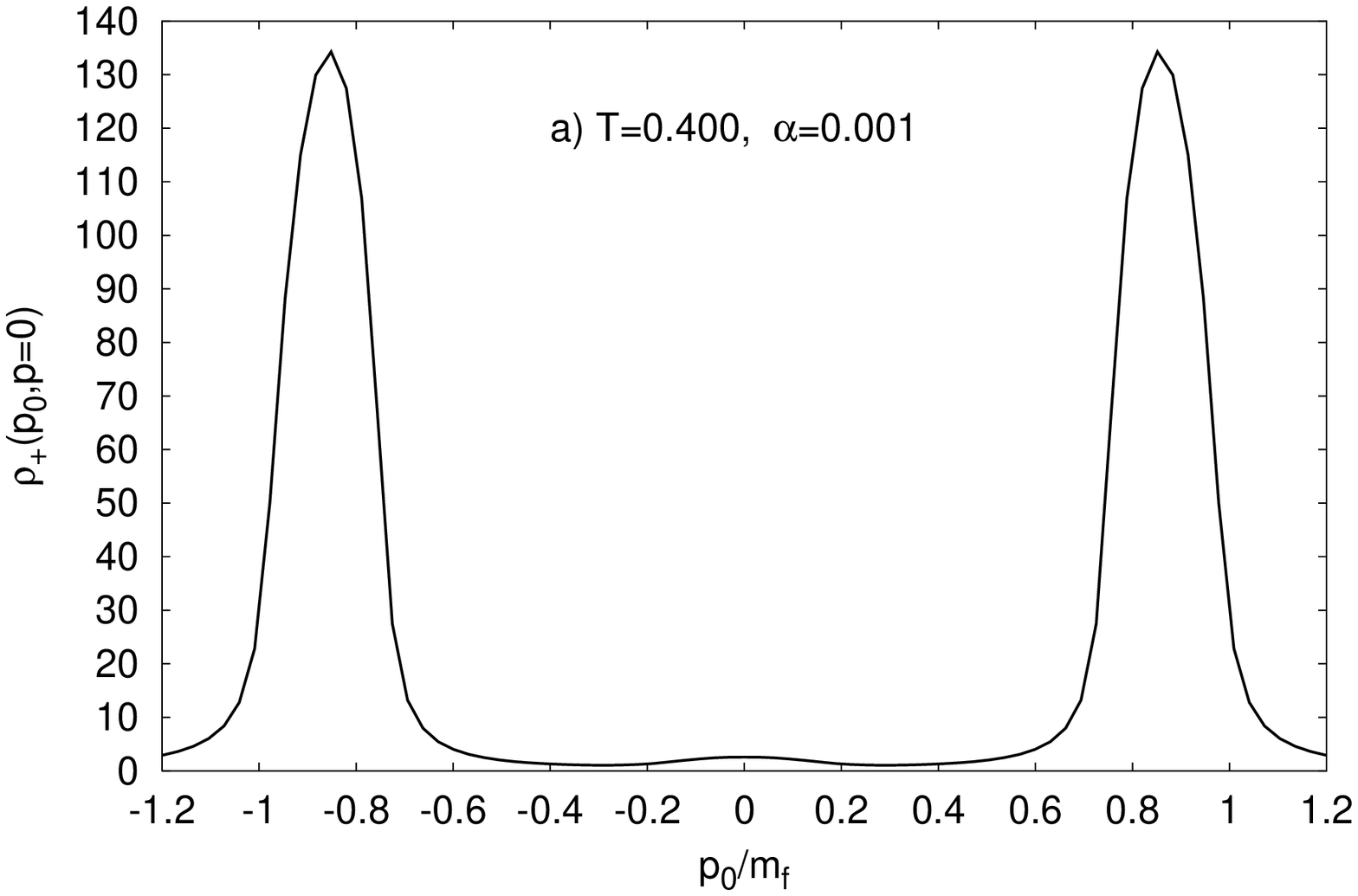}} \vspace{-0.4cm}
  \centerline{\includegraphics[width=7.2cm]{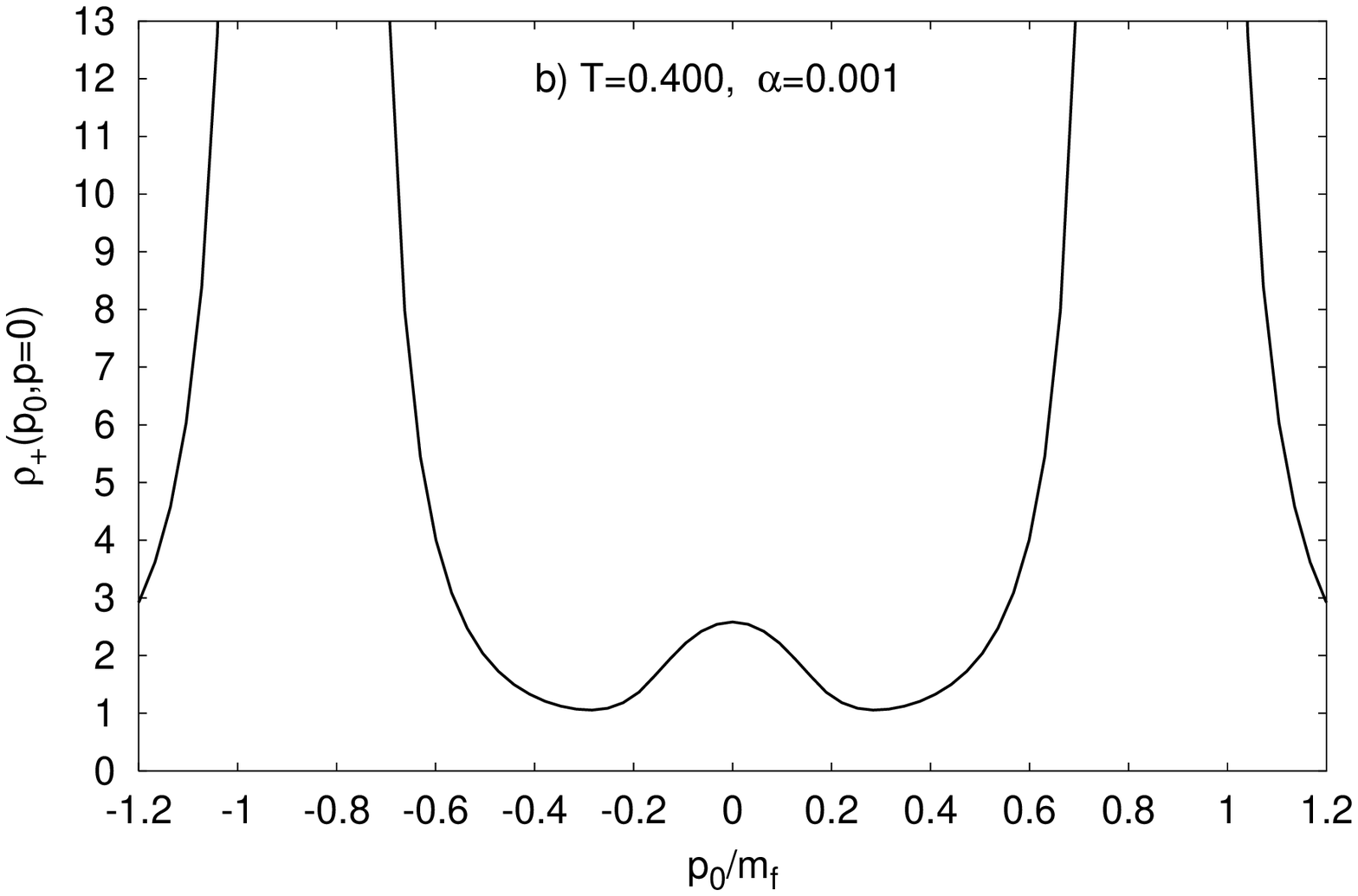}} \vspace{-0.4cm}
\caption{(a) Quasifermion spectral density $\rho_+(p_0,p=0)$ in the small coupling region 
($\alpha=0.001, T=0.4$). \ (b) Quasifermion spectral density $\rho_+(p_0,p=0)$ enlarged around the origin.}
\end{figure}

\section{Imaginary part of the chiral invariant mass function B, or the decay width of the quasifermion} 
Finally let us see the imaginary part of the chiral invariant mass function Im[$B(p_0, p)$]. The decay width of the
quasifermion is extensively studied through the HTL resummed effective perturbation calculation~\cite{Nakk3}, 
giving a gauge invariant result of $O(g^2T \log (1/g))$. However, as is shown above, the quasiparticle exhibits 
an unexpected behavior, such as the vanishing of the thermal mass in the strongly coupled QCD/QED medium, completely different from that 
expected from the HTL resummed effective perturbation analyses. How does the decay width of the quasifermion exhibit 
its property in the corresponding strongly coupled QCD/QED medium?

In Fig.~6 we show the Im[$B(p_0=\omega_+,p=0)$] in the small coupling region, which agrees 
with the HTL resummed calculation~\cite{Nakk3} up to a numerical factor. 
In Fig.~7 we show Im[$B(p_0=\omega_+=0,p=0)$] in the strong coupling region.
\begin{figure}[htbp] 
   \vspace*{-0.5cm}
  \centerline{\includegraphics[width=7.5cm]{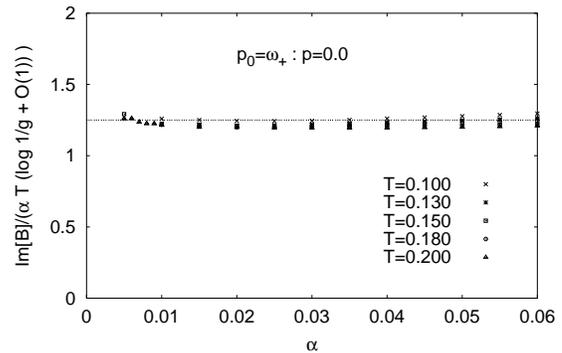}} \vspace{-0.4cm}
  \caption{The Im[$B$] in the small coupling QGP.}
\end{figure}
\begin{figure}[htbp] 
  \vspace*{-0.5cm}
  \centerline{\includegraphics[width=7.5cm]{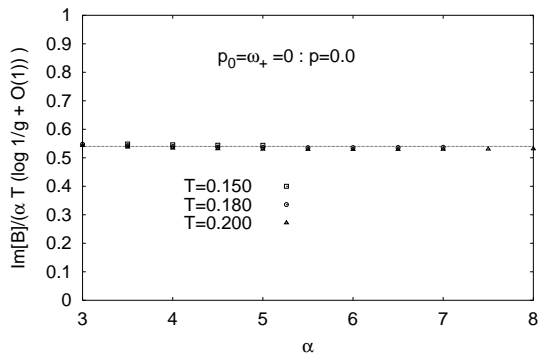}} \vspace{-0.4cm}
  \caption{The Im[$B$] in the strongly coupled QCD/QED medium.}
\end{figure}

As can be seen, even in the strong coupling  region, Im[$B$] or the decay width shows the same behavior of $O(g^2T \log (1/g))$; 
the numerical factor, of course, largely differs from that in the small coupling region. This behavior is again 
not expected, because the quasifermion in the small coupling QGP and the one in the strong 
coupling QGP are totally different; while in the former case the quasifermion has a thermal 
mass of $O(gT)$ and the plasmino branch exists in a fermion dispersion law, in the latter case the thermal mass of the
quasifermion vanishes and the plasmino branch disappears.

\section{Summary} 
In this paper, through the analysis employing the HTL resummed improved ladder DSE, we disclose the fact that 
the property of the thermal quasifermion in the strongly coupled QCD/QED medium largely differs from that expected from the analyses based 
on the HTL resummed effective perturbation calculations. We can summarize our main result as follows.
In the strongly coupled QCD/QED medium, (i) while the thermal mass of the quasifermion vanishes and the quasifermion pole starts to behave as a massless pole,
 (ii) the imaginary part of the chiral invariant mass function, or the decay width of the  quasifermion persists to 
have the behavior of $O(g^2T \log (1/g))$, which is expected in the small coupling and high temperature QCD/QED medium.
   
The first result suggests that in the recently produced sQGP the thermal mass of the quasifermion should vanish or at least 
become significantly lighter compared to the value in the  weakly coupled QGP. And the second result indicates that in 
the sQGP the massless or the ultrasoft fermion pole may not be observed as a massless quasifermion mode due to 
its large imaginary part.
 
Evidence of the existence of a massless, or an ultrasoft mode, at least in a weakly coupled QGP, is also pointed out.

Details of the present DSE analysis and the investigation of the results on the property of the quasifermion will 
be given in a separate paper~\cite{Nakk2}.

\vspace{0.5cm}%
\acknowledgements
We thank the Yukawa Institute for Theoretical Physics at Kyoto University. Discussions during 
the YITP workshop YITP-W-11-14 were useful to complete this work.
Part of the present work is supported by the Nara University Research Fund.
Numerical computation in this work was carried out at the Yukawa Institute Computer Facility.


\begin{references}
\bibitem{foot}
   In solving the DSE, we introduce a momentum cut-off as follows ($\Lambda$ denotes an arbitrary 
   cut-off parameter and plays a role to 
   scale any dimensionful quantity, e.g., $T=0.3$ means $T=0.3 \Lambda$):%
\begin{center}
\begin{tabular}{llcl}
three momentum & ~$p$~ & ~:~ & $p \le \Lambda$  \\
energy         & ~$p_0$~ & ~:~ & $|p_0| \le  a \Lambda$ \  ($a=1 \sim 5$) \\
\end{tabular}
\end{center}
We determine the parameter ``$a$'' so as to get a stable solution for the fermion spectral density. Over 
the range of the temperature and the coupling we study in the present analysis,
we set $a=2$. In fact, the solution is totally stable for $a \ge 2$.
\bibitem{foot2}
The temperature $T$ in the present analysis represents the temperature scaled by the momentum 
cutoff  $\Lambda$, and it does not denote the real temperature itself. With the temperature thus 
defined, we can determine the phase-boundary curve in the two-dimensional ($\alpha$, $T$) plane, 
separating the chiral symmetry broken/restored phase~\cite{Nakk}. If we measure the 
temperature $T$ relative to the critical temperature $T_c$, high or low temperature 
has a definite meaning.
\bibitem{foot3}
  The ratio $\omega_+(p=0)/T$ is not necessarily a constant, and the $T$ dependence 
at small $T$ obtained in our analysis is a consequence of the nonperturbative DSE
analysis. This is because the additional dimensionful parameter, such as the
regularization (or the cutoff) scale or the renormalization scale comes into the theory 
through the regularization and/or the renormalization of massless thermal QCD/QED. 
In our case the cutoff scale $\Lambda$ is introduced into the theory.  
The thermal mass, in fact, has a log~$T$ dependence 
in the effective perturbation calculation: see, e.g., Rebhan's lecture in Ref.~\cite{Rebhan}.
\bibitem{foot4}
The temperature-dependent behavior of the thermal mass at extremely high 
temperature ($T \simeq 1$ in our present analysis) may exhibit a simulation
or numerical  artifact, and 
the $T$ dependence at $T \simeq 1$ 
needs further careful analyses.
\end{references}
\end{document}